\documentclass[aip,jcp,graphicx,reprint,longbibliography,noeprint,citeautoscript]{revtex4-1}

\usepackage[pdftex]{graphics}
\usepackage{amssymb,amsmath}
\usepackage{bm}
\usepackage{siunitx}

\newcommand{\ii}{\mathrm{i}}
\newcommand{\kBT}{k_\mathrm{B}T}

\newcommand{\au}{\text{a.u.}}
 
\renewcommand{\Re}{\operatorname{Re}}
\renewcommand{\Im}{\operatorname{Im}}
\newcommand{\erfc}{\operatorname{erfc}}

\usepackage{dcolumn} 
\newcolumntype{d}[1]{D{.}{.}{#1}} 

\makeatletter
\let\@fnsymbol\@fnsymbol@latex
\@booleanfalse\altaffilletter@sw
\makeatother

\begin{document}

\title{On decoherence in surface hopping: the nonadiabaticity threshold}

\author{Johan E. Runeson}
\email{johan.runeson@physik.uni-freiburg.de}
\affiliation{Institute of Physics, University of Freiburg, 79104 Freiburg, Germany}

\begin{abstract}
This work presents a strategy to efficiently and safely account for decoherence in the fewest switches surface hopping method. Standard decoherence corrections often lead to too strong coherence suppression.
A simple and general solution to this problem is to restrict decoherence to regions of low nonadiabaticity measured by the dimensionless Massey parameter. The same threshold values are suitable for a variety of systems, regardless of their size and absolute energy scale. When restricted to uncoupled regions, a Gaussian overlap decoherence correction consistently leads to more accurate populations than using no correction. The article also examines under what circumstances it is appropriate to decohere instantaneously.
\end{abstract}

\maketitle

\section{Introduction}
In the field of mixed quantum-classical dynamics, few topics have drawn as much attention in the last decades as the problem of decoherence in surface hopping. Tully's fewest switches surface hopping (FSSH)\cite{tully1990hopping} is a leading method for excited-state dynamics in molecules and materials, with extensive impact on contemporary physics and chemistry.\cite{Barbatti2018review} The method is widely available in public software and enables -- at least  qualitatively -- a simple understanding of complex dynamical processes in terms of classical trajectories hopping between discrete quantum states. However, a crucial flaw of the method remains its lack of proper description of decoherence.\cite{Subotnik2016review} Much effort has been made to correct for this error, but since the corrections introduce more or less uncontrolled approximations, their range of validity is generally unknown.

In a fully quantum simulation, one has direct access to the correct inter-surface coherence through the overlaps of the nuclear wavefunctions on different surfaces.
Due to the exponential scaling of quantum methods with system size, such simulations are only possible for very simple model systems. The purpose of FSSH (and other mixed quantum-classical) simulations is to avoid explicit propagation of the full wavefunction, but the massive reduction in computational effort comes at the expense of losing access to the overlap. Hence, it has been necessary to account for decoherence using \emph{ad hoc} methods. In the 1990s, 
Hammes-Schiffer and co-workers suggested collapsing the wavefunction whenever the system enters a region where the nonadiabatic coupling vector has a small magnitude.\cite{Hammes-Schiffer1994, Fang1999improvement} This method unfortunately relies on system-dependent thresholds, but is nevertheless an important inspiration for the work of the present paper. In the same decade, Schwartz, Bittner, Prezhdo, and Rossky estimated the rate of decoherence from the decay of overlap between frozen Gaussians.\cite{Schwartz1996decoherence} They obtained a rate proportional to the force difference between the two surfaces. Through different assumptions, other authors have arrived at a variety of schemes based on Gaussian overlaps.\cite{Neria1993,Jasper2005decoherence,Granucci2010,shenvi2011STSH} Later on, Subotnik and co-workers developed an ``augmented'' method (A-FSSH) in which one estimates the overlap decay rate by propagating auxiliary variables for the wavepacket moments.\cite{subotnik2011AFSSH,Jain2016efficient} An alternative class of decoherence corrections, pioneered by Truhlar and co-workers, estimate the decoherence rate based on the energy difference.\cite{Hack2001,Zhu2004selfconsistent,Zhu2004csdm,Zhu2005jtct} Granucci and Persico suggested a simplified version of their formula, known as the energy-based decoherence correction (EDC).\cite{granucci2007} Thanks to its simplicity, the EDC has become widely popular,\cite{Barbatti2018review} although its results are known to be sensitive to the choice of parameter values.\cite{Nelson2013} Lately, there has also been a growing interest in more sophisticated (but computationally demanding) coupled-trajectory approaches.\cite{Pieroni2021coupled,Pieroni2024coupled,Xie2025ctsh}
Apart from the decoherence schemes just mentioned, many more have been described in recent reviews.\cite{Wang2016recentprogress,Barbatti2018review,Carof2019mobility, Nelson2020review,Wang2020review,VillasecoArribas2023review,Shu2023decoherence}

One way to mitigate the problem of overcoherence is to switch hopping criterion. The mapping approach to surface hopping (MASH) provides an alternative framework based on deterministic hops to the most populated surface.\cite{Mannouch2023mash,Runeson2023mash,Runeson2025lhc} But just as in FSSH, there are multiple ways to measure populations also in MASH, which are not always guaranteed to be consistent with each other. Although MASH is usually more accurate than FSSH without a decoherence correction, it still does not eliminate the need for one.\cite{Lawrence2023mash} Since MASH and FSSH rely on rather different decoherence algorithms, the present article will focus solely on FSSH, but its conclusions are expected to be relevant also to MASH.

An underrated problem when using decoherence corrections is that the ``corrected'' simulation is not guaranteed to be more accurate than the uncorrected simulation. If the crudely estimated decoherence rate happens to be too large, one may even destroy the correct coherent evolution of the quantum subsystem. The main purpose of the present paper is to propose a method to avoid this problem. In general, the decoherence rate is most difficult to calculate accurately in regions of strong nonadiabatic coupling, where the potentials are strongly anharmonic. By restricting decoherence to regions of low nonadiabaticity, measured in terms of the dimensionless Massey parameter, the paper shows that one can protect the ``correct'' coherence and achieve consistent improvements over uncorrected FSSH across very different types of systems.
In the ``safe'' regions, the paper suggests using a refined formula for the frozen-Gaussian decoherence rate. This formula originates from the time-dependent overlap of Gaussians evolving on different surfaces, and does not require any auxiliary trajectories. Apart from the force difference, the formula also depends on the momentum difference that would arise from rescaling at a hop. The paper also considers an approximate alternative to this formula where decoherence is treated as instantaneous.


The article begins with a summary of the (decoherence-free) FSSH method in section~\ref{sec:FSSH}. It then narrows down on the topic decoherence in \ref{sec:decoherence}, where it revisits the calculation of overlap decay rates from Gaussian overlaps, shows why a nonadiabaticity threshold is necessary, and discusses important objections against the EDC. The methods are illustrated by an example calculation for Tully's famous extended coupling model. In section~\ref{sec:results}, the methods are compared for a suite of benchmark systems with varying size and absolute energy scales. Finally, the conclusions are presented in section~\ref{sec:conclusions}.

\section{Fewest switches surface hopping}\label{sec:FSSH}
The following section summarises Tully's fewest switches surface hopping method without decoherence corrections.\cite{tully1990hopping} Special attention is given to the algorithmic choices that concern velocity rescaling, observable measurement, and initial conditions. 

The starting point for mixed-quantum classical dynamics is a Hamiltonian of the form
\begin{equation}
    H = \sum_j \frac{p_j^2}{2m_j} + V(q).
\end{equation}
Here, $p$ and $q$ are the degrees of freedom that will be treated classically (usually the nuclear/vibrational modes) and
\begin{equation}
    V(q) = \sum_{nm}V_{nm}(q)|n\rangle\langle m|
\end{equation}
is the potential operator of the quantum subsystem (usually the space of electronic excitations). In this expression, the basis is assumed to be diabatic (i.e., $\nabla_j|n\rangle = 0$). One can always convert $V(q)$ to its local eigenbasis, called the adiabatic basis,
\begin{equation}
    V(q) = \sum_a V_a(q)|a(q)\rangle\langle a(q)|.
\end{equation}
Throughout this paper, the indices $n,m$ refer to a given diabatic basis, while the indices $a,b$ refer to the adiabatic basis. The transformation matrix between the two bases is denoted by $U_{na}=\langle n|a(q)\rangle$.

The goal of mixed quantum-classical dynamics is to combine the evolution of the quantum subsystem with a suitable classical approximation for the evolution of $p$ and $q$. 
To this end, we describe the state of the quantum subsystem by a wavefunction
\begin{equation}
    |\psi\rangle = \sum_n c_n |n\rangle = \sum_a c_a|a(q)\rangle,
\end{equation}
which follows the time-dependent Schr\"{o}dinger equation along the trajectory, 
\begin{equation}\label{eq:TDSE}
    |\dot{\psi}\rangle = -\frac{\ii}{\hbar}V(q)|\psi\rangle,
\end{equation}
whereas the classical variables follow Hamilton's equation of motion on some effective potential $V_\text{eff}(q)$,
\begin{align}
    \dot{q}_j &= p_j/m_j \\
    \dot{p}_j &= -\nabla_j V_\text{eff}(q).
\end{align}
There are many possibilities to choose the effective potential. For example, the Ehrenfest approach is to use the expectation value of the potential operator, $V_\text{eff}(q) = \langle \psi|V(q)|\psi\rangle$. This choice is problematic when passing through a local nonadiabatic crossing, because the trajectory would emerge on a superposition of surfaces, whereas the true quantum wavefunction (obtained by including all variables in the quantum subsystem) splits into one component on each state. A better strategy to reproduce such a bifurcation event is to only evolve on eigenstates, as opposed to superpositions of eigenstates. 
We say that a trajectory has ``active state'' $a$ when it evolves on 
\begin{equation}\label{eq:Veff}
    V_\text{eff}(q) = \langle a(q)|V(q)|a(q)\rangle.
\end{equation}
The main idea of Tully's surface hopping is to, as time progresses, switch surfaces in such a way that the fraction of trajectories with active state $a$ matches the wavefunction population $P_a=|c_a|^2$. 

Consider an example with two coupled states, $a$ and $b$, where initially $P_a=0.6$ and $P_{b}=0.4$, and at a later time $P_a=P_{b}=0.5$. The fewest switches necessary to match the change in populations would be to let one in six of the trajectories on $a$ switch to $b$, and none of the trajectories on $b$ switch to $a$. Any additional hops would be redundant. More generally (but still for two states), the rate of population transfer from $a$ to $b$ is $\dot{P}_b/P_a$. To find an expression for the time derivative of the population, $\dot{P}_b = 2\Re(c_b^* \dot{c}_b)$, we need the TDSE for the adiabatic coefficients. Starting from the equation of motion for the diabatic coefficients,
\begin{equation}\label{eq:cdia}
    \dot{c}_n = -\frac{\ii}{\hbar} \sum_m V_{nm}(q)c_m
\end{equation}
one can easily derive the equation of motion in the adiabatic basis by using the product rule on $c_n=\sum_a U_{na}c_a$, giving
\begin{equation}\label{eq:cad}
    \dot{c}_a = -\frac{\ii}{\hbar} V_a c_a - \sum_b T_{ab} c_b,
\end{equation}
where the time-derivative coupling $T=U^{-1}\dot{U}$ has elements
\begin{equation}
    T_{ab} = \langle a(q)|\frac{d}{dt}b(q)\rangle 
\end{equation}
In this way, one arrives at 
\begin{equation} \label{eq:Pbdot_over_Pa}
    \frac{\dot{P}_b}{P_a} = -\frac{2 \Re(c_b^* T_{ba} c_a)}{|c_a|^2} = \frac{2 \Re(c_a^* T_{ab} c_b)}{|c_a|^2},
\end{equation}
where in the last equality we used $T_{ba}=-T_{ab}^*$.
The FSSH approach is therefore to, in each timestep $\Delta t$, hop from $a$ to $b$ with probability
\begin{equation}\label{eq:Phop}
    P_{a\to b} = \max\left[0, \frac{2 \Re(c_a^* T_{ab} c_b) }{|c_a|^2}\Delta t \right].
\end{equation}
The lower bound of zero means that we do not hop to states for which the population is decreasing. 
If there are more than two coupled states in the quantum subsystem, one uses the two-state formula to calculate the hopping probability to each non-active state, and selects the new state $b'$ that fulfils
\begin{equation}\label{eq:xi}
    \sum_{b=1}^{b'-1} P_{a\to b} < \xi < \sum_{b=1}^{b'} P_{a\to b},
\end{equation}
where $\xi$ is a random number sampled uniformly from $[0,1)$. (The time step is assumed to be small enough to contain the cumulative probabilities within this interval.)

Before we continue, three practical considerations are worth mentioning. 
First, if one has access to a diabatic basis (which we do for all systems cases considered in this paper), a stable way to integrate the equation of motion for the wavefunction (assuming fixed $q$) is 
\begin{equation}\label{eq:c_integrator}
    {c}_n(\Delta t) = \sum_a U_{na}(q) e^{-\ii V_a(q)\Delta t/\hbar} \sum_m U_{am}(q) c_m(0).
\end{equation}
This procedure is preferable to integrating Eq.~\eqref{eq:cad}, because the diabatic basis vectors $|n\rangle$ are constant in time, whereas $T_{ab}$ may change rapidly. Second, one can calculate $T_{ab}$ using the chain rule,
\begin{equation} \label{eq:Tab}
    T_{ab} = \sum_j d_{ab}^j \dot{q}_j
\end{equation}
where 
\begin{equation}
    d_{ab}^j = \langle a(q)|\nabla_j b(q)\rangle 
\end{equation}
are the elements of the nonadiabatic coupling vector (NACV). Using the Hellman--Feynman theorem, the latter is obtained from the potential and its gradient as
\begin{equation}
    d_{ab}^j = \frac{\langle a(q)|\nabla_j V(q)|b(q)\rangle}{V_b-V_a}.
\end{equation}
If the NACV varies rapidly, one may alternatively calculate $T_{ab}$ using numerical derivatives,\cite{Hammes-Schiffer1994} 
\begin{equation} \label{eq:Tab2}
    T_{ab}\left(t+\frac{\Delta t}{2}\right) \approx \frac{1}{2\Delta t}[\langle a(q_t)|b(q_{t+\Delta t})\rangle - \langle a(q_{t+\Delta t})|b(q_t)\rangle],
\end{equation}
which does not involve the NACV. For the systems considered in the present work, Eq.~\eqref{eq:Tab} was found to be preferable because it reached convergence with a larger time step than Eq.~\eqref{eq:Tab2}.
Finally, since adiabatic state vectors do not have a well-defined global phase, $T_{ab}$ and $d_{ab}$ should be computed using phase-corrected state vectors. A simple phase correction suggested by Akimov is\cite{Akimov2018} 
\begin{equation}
    |a(q_{t+\Delta t})\rangle \leftarrow |a(q_{t+\Delta t})\rangle \frac{\langle a(q_{t+\Delta t}|a(q_t)\rangle}{|\langle a(q_{t+\Delta t}|a(q_t)\rangle|}.
\end{equation}
In the present work, the phase correction did not lead to any notable changes [because the phase cancels out automatically in Eqs.~\eqref{eq:Veff},~\eqref{eq:Phop} and~\eqref{eq:c_integrator}], but it is necessary to uniquely define the NACV direction used in the momentum rescaling.


\subsection{Momentum rescaling}\label{sec:rescaling}
When the trajectory hops from $a$ to $b$, it is physically justified to adjust its kinetic energy in order to conserve the total energy. A concensus has emerged that the appropriate direction for momentum rescaling is along the NACV ($d_{ab}$). In case the kinetic energy is insufficient to overcome the energy barrier (a so-called frustrated hop), one instead reflects the momentum projection along the same direction and leaves the active state at $a$. This procedure (rescaling and reversal along the NACV) is precisely what Tully originally proposed,\cite{tully1990hopping,Hammes-Schiffer1994} it is justified on theoretical grounds,\cite{Herman1984,Tully1991,Coker1995} and a large body of empirical evidence supports it over alternative procedures.\cite{Hack1999,Plasser2019rhenium,Mannouch2024kelly}
In particular, rescaling along the NACV has been empirically found to approximately recover detailed balance (correct long-time electronic populations).\cite{Parandekar2005mixed,Kaeb2006,schmidt2008SH,Carof2017FOB-SH} Rescaling the entire momentum easily breaks detailed balance, because it allows crossing energy gaps on the order of the total kinetic energy (an extensive quantity). In contrast, rescaling along a given component of the momentum only allows crossing energy gaps on the order of $\kBT$ (an intensive quantity).\cite{Plasser2019rhenium} Any given direction would fulfil this criterion, but the particular direction that is relevant for driving population flow is the one that appears in Eq.~\eqref{eq:Tab}, namely the NACV.

It is helpful to think of the system impacting on a potential wall analogous to how a light ray crosses between media with different refractive indices. In both cases, it is the momentum orthogonal to the wall (wave vector orthogonal to the boundary) that needs to be adjusted. In the recent MASH method, where trajectories are deterministic, it emerges directly from the form of the effective potential (for two states) that this direction is indeed $d_{ab}$.\cite{Mannouch2023mash} For multi-state MASH, where trajectories hop to the most populated surface, the rescaling direction involves a linear combination of different NACVs to account for population flows involving third states.\cite{Runeson2023mash} The latter direction would not be suitable for FSSH, because the hopping probability in Eq.~\eqref{eq:Phop} does not involve third states.

To abbreviate the notation, let $d=d_{ab}$ where $a$ is the current active state and $b$ is the new state.
In case of non-uniform masses, it is easiest to work in mass-scaled variables and rescale/reflect $p/\sqrt{m}$ along $d/\sqrt{m}$.\cite{Lawrence2024cyclobutanone} The projection of $p$ parallel to the direction of rescaling is then
\begin{equation}
    p^{\rm par} = \frac{\sum_j d_j p_j / m_j}{\sum_j d_j^2 / m_j} d
\end{equation}
and has the associated kinetic energy
\begin{equation}
    E_{\rm kin}^{\rm par} = \sum_j \frac{(p_{j}^{\rm par})^2}{2m_j}.
\end{equation}
For a potential step $\Delta V = V_b-V_a$, the adjusted momentum after the hop is
\begin{equation}\label{eq:prescale}
    p' = \begin{cases}  \displaystyle
    p + p^{\rm par} \left(\sqrt{1-\frac{\Delta V}{E_{\rm kin}^{\rm par}}} - 1\right) & \text{if } \Delta V \leq E_{\rm kin}^{\rm par} \\
    p - 2p^{\rm par} & \text{if }\Delta V > E_{\rm kin}^{\rm par}.
    \end{cases}
\end{equation}
This formula gives identical results to the original rescaling formula by Hammes-Schiffer and Tully.\cite{Hammes-Schiffer1994}

\subsection{Observables}
Another source of variety between FSSH implementations is how to measure electronic populations and coherences. On the one hand, one can simply calculate the entire reduced density matrix from the wavefunction:
\begin{equation}\label{eq:rhoP}
    \rho_{\rm P} = |\psi\rangle\langle \psi| = \sum_{ab} c_a^* c_b|a\rangle \langle b|.
\end{equation}
Then the diagonal elements in the adiabatic basis are the adiabatic wavefunction populations $P_a=|c_a|^2$. On the other hand, one could equally well measure the diagonal elements by the active states. Denoting the active state by $a'$, the corresponding ``mixed'' estimator is\cite{Landry2013interpretation}
\begin{equation}\label{eq:rhoPi}
    \rho_\Pi = \sum_{a}\delta_{aa'}|a\rangle\langle a| + \sum_a \sum_{b\neq a} c_a^* c_b|a\rangle \langle b|.
\end{equation}
The diagonal elements of $\langle \rho_\Pi\rangle$ (where $\langle \cdots\rangle$ denotes an average over the swarm of trajectories) is the fraction of trajectories on each adiabatic state. Since $\rho_\Pi$ accounts for frustrated hops, it is usually considered as a more accurate way to measure population transfer than $\rho_{\rm P}$. But any discrepancy between the two is an indication of internal inconsistency, and ideally, one would like the two to be as close as possible (how to achieve this will be addressed in section~\ref{sec:decoherence} on decoherence). 

Note that both $\rho_{\rm P}$ and $\rho_{\Pi}$ can be transformed to any basis, so each of them allows measuring diabatic as well as adiabatic observables. Explicitly, the diabatic populations are
\begin{equation}\label{eq:P_n}
    P_n = |c_n|^2
\end{equation}
for the wavefunction measure and
\begin{equation} \label{eq:Pi_n}
    \Pi_n = \sum_a U_{na'}^2 + \sum_a \sum_{b\neq a}c_a^*  U_{na}^* U_{nb}c_b
\end{equation}
for the ``standard'' mixed measure (again, $a'$ denotes the active state).
In principle, $\Pi_n$ can be less than zero or greater than one, because there is no guarantee that $\rho_{\Pi}$ is positive definite (in practice, this situation is rare).
Previous work\cite{Landry2013interpretation} have also considered a third measure that only includes the first term in $\Pi_n$, but this measure leads to incorrect initial values for diabatic populations and will not be considered in this paper.


\subsection{Initial conditions}
When starting in an adiabatic state $|a\rangle$, one simply sets $|\psi\rangle = |a\rangle$ and assigns $a$ as the active state.
When starting in a diabatic state $|n\rangle$, one sets $|\psi\rangle = |n\rangle$ and samples the initial active state $a$ from a set of probabilities $P_a = |\langle a|\psi\rangle|^2 = |U_{na}|^2$. 
A slight variation present in the literature is to first sample $a$ and subsequently reset $|\psi\rangle$ to $|a\rangle$. The latter step is unnecessary and unjustified, since it makes the initial diabatic populations different from what they were intended to be.

\section{Decoherence}\label{sec:decoherence}
This section illustrates the problem of overcoherence with example calculations for Tully's extended coupling model. First, I show that a decoherence rate based on Gaussian overlaps leads to too strong coherence suppression. Then I show how to solve this problem by introducing a nonadiabaticity threshold. Finally, I discuss why this approach is preferable to the commonly used energy-based decoherence correction.

To introduce the problem of overcoherence, consider an initial wavepacket on $|a\rangle$ passing through a local nonadiabatic crossing with another state $|b\rangle$. In a fully quantum calculation, the wavefunction splits into transmitted and/or reflected components on each state. In the FSSH algorithm, each trajectory emerges on a physical eigenstate, and the fraction of trajectories along each channel closely approximates the scattering probabilities of the quantum wavepacket. However, the FSSH wavefunction remains a coherent superposition of eigenstates. In other words, the coherence $\langle a|\psi\rangle\langle \psi|b\rangle=c_a^* c_b$ generally remains non-zero 
even when the coherence of the quantum calculation vanishes. 

\begin{figure*}
    \includegraphics{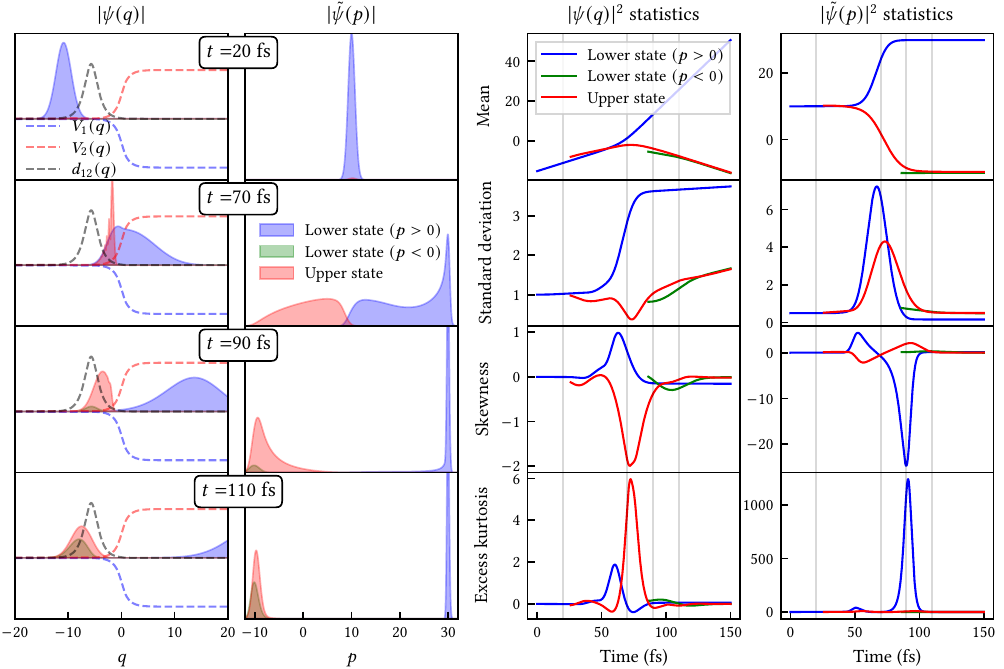}
    \caption{Quantum propagation in Tully's extended coupling model. Columns 1-2: snapshots of the wavepacket in its $q$ and $p$ representations, decomposed into forward and backward components on the lower adiabat and the total component on the upper adiabat. The wavefunction magnitude is shown in place of its square to make small and large components visible on the same scale. Columns 3-4: time evolution of descriptors of the $q$ and $p$ densities. Vertical lines indicate the times of the snapshots. In all columns, $q$ and $p$ are measured in atomic units.}
    \label{fig:tully3WP}
\end{figure*}

The lack of decoherence in the FSSH method can have severe consequences for the accuracy of the results. For systems with multiple crossings, the trajectories will typically enter the second crossing with too much ``memory'' intact from the first crossing. An illustrating example is Tully's extended coupling model (also known as ``Tully's third system''). The diabatic model potential is\cite{tully1990hopping}
\begin{align}
    V_{11} = A, \quad V_{22} = -A \nonumber \\
    V_{12} = \begin{cases} B e^{+Cq} & q<0 \\ 
    B(2-e^{-Cq}) & q>0 \end{cases} \label{eq:tully3}
\end{align}
where $A=6\times 10^{-4}$, $B=0.1$, $C=0.9$ and $m=2000$ in atomic units.
The adiabatic potentials and NACV are shown in the left column of Fig.~\ref{fig:tully3WP} along with snapshots of the wavepacket evolution (using the split-operator method). The initial wavepacket starts on the lower state and is of the form 
\begin{equation}\label{eq:WPgauss}
    \psi_0(q) = \left(\frac{\gamma}{\pi}\right)^{1/4}\exp\left[-\frac{\gamma}{2}(q-q_0)^2 + i\frac{p_0}{\hbar}(q-q_0)\right]
\end{equation}
with $\gamma=0.5$, $q_0=-15$, and $p_0=10$. In FSSH, the initial values for $q$ and $p$ are sampled from its Wigner distribution 
\begin{equation}
    \rho(q,p) = \frac{1}{\pi\hbar} \exp\left[-\gamma(q-q_0)^2 - \frac{1}{\gamma\hbar^2}(p-p_0)^2 \right].
\end{equation}
After the first crossing, the wavepacket splits into one component on each surface. The upper-state component subsequently reflects and passes through the crossing a second time. Without accounting for decoherence, FSSH only describes the first crossing correctly, as shown in the top row of Fig.~\ref{fig:tully3}. After the second crossing, there is notable inconsistency between the active-surface population measure $\Pi$ (solid lines) and the wavefunction population measure $P$ (dashed lines) -- and neither agrees with the quantum result.

For this reason, it is necessary to supplement the FSSH method with a decoherence correction. However, since one does not generally know the decoherence rate of the quantum wavepacket (after all, the aim of FSSH is to avoid propagating the wavepacket) the best one can hope for is a more or less crude approximation of the decoherence rate.

\subsection{Gaussian overlaps}\label{sec:Gaussian}
Clearly, the correct decoherence rate depends on the form of the quantum wavefunction. Following a long tradition in semiclassical dynamics,\cite{Heller1975,Herman1984gaussian} we shall focus on the Gaussian wavepacket in Eq.~\eqref{eq:WPgauss}.

Consider a process where the initial wavefunction of the coupled system is $|\psi_0\rangle \otimes |a\rangle$.
After passing through a crossing between states $a$ and $b$, the wavefunction bifurcates into the form $|\psi_a\rangle\otimes |a\rangle + |\psi_b\rangle\otimes|b\rangle$.
If one assumes that $\psi_a$ and $\psi_b$ are also Gaussian and have the same width parameter as $\psi_0$, then it is well known that
\begin{equation}
    |\langle \psi_a|\psi_b\rangle|^2 = \exp\left[-\frac{\gamma}{2}\Delta q^2 - \frac{1}{2\gamma\hbar^2}\Delta p^2\right]
\end{equation}
where $\Delta q=q_b-q_a$ and $\Delta p=p_b-p_a$ are the distance between the centres in coordinate and momentum space. This result was the starting point for a decoherence correction by Jasper and Truhlar, for the force-based correction suggested by Schwartz and coworkers,\cite{Schwartz1996decoherence} as well as for the overlap decoherence correction (ODC) by Granucci, Persico and Zoccante.\cite{Granucci2010}
In the following, I derive a decoherence time that differs slightly in the details from each of these previous works, but nevertheless is based on the same physical ideas.

To estimate the rate of decoherence, we can use the following thought experiment.
At each time step, our trajectory either hops or remains at the current active surface.
How quickly does a Gaussian on a hopping trajectory decohere with a Gaussian on a trajectory that does not hop? 
We want to answer this question using information available from only one trajectory (to avoid running auxiliary trajectories) at one point in time (to avoid storing the history of the trajectory). 
The two Gaussians emerge from the same place, so $\Delta q(0)=0$, but they will generally have a non-zero $\Delta p(0)\equiv\Delta p_{\rm hop}$ due to momentum rescaling along the NACV [Eq.~\eqref{eq:prescale}]. (Side note: the ODC involves rescaling of the entire momentum.\cite{Granucci2010}) From Heller,\cite{Heller1975} we know that distances between two Gaussian centres (in coordinate and momentum space) follow the equations of motion
\begin{subequations}
    \begin{align}
    \Delta \dot{q} &= \Delta p/m \\
    \Delta \dot{p} &=  F_b-F_a \equiv \Delta F
    \end{align}
\end{subequations}
The width parameter $\gamma$ will generally also be time-dependent except when the potentials are harmonic. 
To simplify the following discussion, assume that $\gamma$ as well as $\Delta F$ are constant on the timescale relevant for decoherence. Then the overlap decays as
\begin{equation}
    |\langle \psi_a|\psi_b\rangle|^2 = \exp(-k_1t-k_2 t^2) \equiv D_\text{Gauss}(t),
\end{equation}
where
\begin{subequations}\label{eq:overlapAB}
\begin{align}
    k_1 &= \frac{1}{\gamma\hbar^2}\Delta p_{\rm hop} \cdot \Delta F \label{eq:overlapB} \\
    k_2 &= \frac{\gamma}{2}\left(\frac{\Delta p_{\rm hop}}{m}\right)^2 + \frac{1}{2\gamma \hbar^2}\Delta F^2.  \label{eq:overlapA}
\end{align}
\end{subequations}
Note that $k_1$ and $k_2$ are computed separately for each non-active state, because each state leads to different $\Delta p_{\rm hop}$ and $\Delta F$.
Since $\Delta p_{\rm hop}$ is defined as the momentum difference arising from a hop, frustrated hops are treated in the same way as in Eq.~\eqref{eq:prescale}, i.e., with reflection along the NACV ($\Delta p_{\rm hop}=-2p^{\rm par}$) and with the same force as the non-hopping trajectory ($\Delta F=0$), meaning that $k_1=0$ in this case. In the energetically allowed case, $k_1$ may be positive or negative, but $k_2$ is always positive, so $D_\text{Gauss}(t)$ will always vanish as $t\to\infty$. For multi-dimensional problems with diagonal $\gamma$ and $m$ matrices, Eq.~\eqref{eq:overlapAB} is to be understood as
\begin{subequations}
\begin{align}
    k_1 &= \sum_j \frac{(\Delta p_{\rm hop})_j (\Delta F)_j}{\gamma_j\hbar^2} \\
    k_2 &= \sum_j \left[\frac{\gamma_j(\Delta p_{\rm hop})_j^2}{2m_j^2} + \frac{(\Delta F)_j^2}{2\gamma_j \hbar^2}\right].
\end{align}
\end{subequations}

\begin{figure*}
    \includegraphics{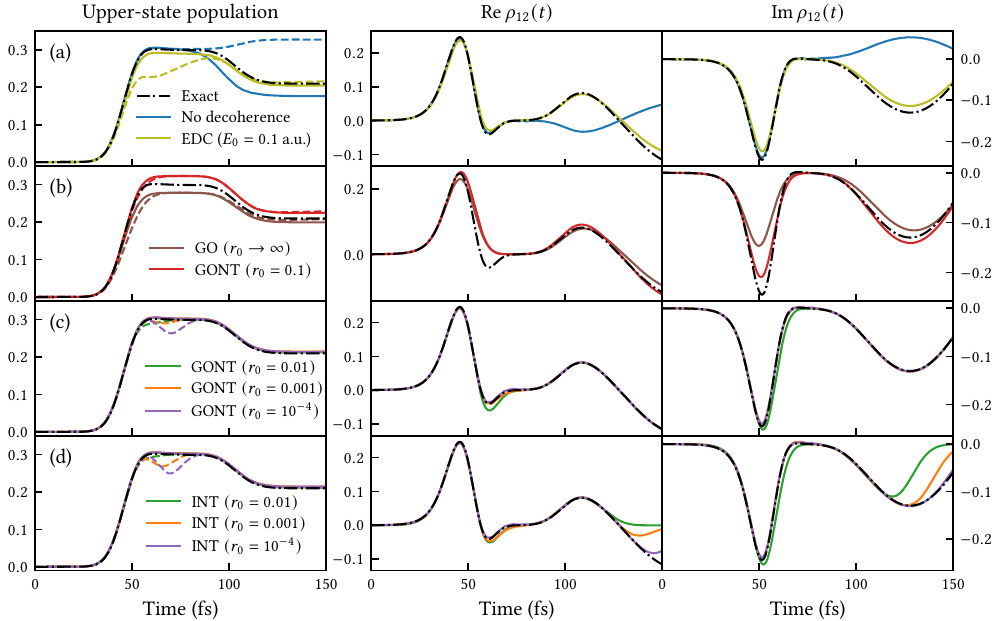}
    \caption{FSSH results for Tully's extended coupling model. The initial conditions are the same as in Fig.~\ref{fig:tully3WP}. Adiabatic populations (left column) are measured with the active surface ($\Pi_2$, solid lines) as well as the wavefunction ($P_2$, dashed lines). Adiabatic coherences (mid and right columns) are measured with the wavefunction only. (a) FSSH without decoherence and with the energy-based decoherence correction. (b) Gaussian overlap decoherence applied universally ($r_0\to\infty$) and with a loose nonadiabaticity threshold ($r_0=0.1$). (c) Gaussian overlap decoherence with the tighter nonadiabaticity thresholds $r_0=0.01$, 0.001, and $10^{-4}$. (d) Instantaneous decoherence with the same nonadiabaticity thresholds.}\label{fig:tully3}
\end{figure*}

The overlap function $D_\text{Gauss}(t)$ may be considered as intermediate between purely Gaussian and purely exponential decay, which is in line with the expected behaviour.\cite{Shu2023decoherence} Due to different initial conditions, Jasper and Truhlar obtained a purely exponential decay,\cite{Jasper2005decoherence} while Schwartz and co-workers obtained a pure Gaussian including only the second term in Eq.~\eqref{eq:overlapA}.\cite{Schwartz1996decoherence} The same term does not appear in the ODC method.\cite{Granucci2010}

Our next step is to make the FSSH overlap function 
\begin{equation}
    D_\text{FSSH}(t) = |c_a(t)|^2|c_b(t)|^2
\end{equation}
decay with the same timescale as $D_\text{Gauss}(t)$. Consider a pure initial coherence with $|c_a(0)|=|c_b(0)|=\frac{1}{\sqrt{2}}$ and arbitrary phases, and let $a$ denote the active state. Then a simple linear damping ansatz, $\dot{c}_b=-(1/\tau)c_b$ (which was suggested by Tully\cite{tully1990hopping} and has become the standard procedure in most decoherence corrections\cite{Barbatti2018review}), combined with the constraint $|c_a|^2+|c_b|^2=1$ leads to
\begin{equation}\label{eq:Dfssh}
D_\text{FSSH}(t)=\tfrac{1}{2}e^{-2t/\tau} (1-\tfrac{1}{2}e^{-2t/\tau}).
\end{equation}
Due to their different functional forms, one should not expect $D_\text{FSSH}(t)$ to reproduce the details of $D_\text{Gauss}(t)$. To define an ``optimal'' damping rate $1/\tau$, we impose that the two overlap functions decay with the same expected time
\begin{equation}
    \langle t \rangle = \frac{\int dt \, t\, D(t)}{\int dt \, D(t)}.
\end{equation}
The integrals can be done analytically for both cases:
\begin{subequations}
\begin{align} 
    \langle t\rangle_{\rm Gauss} &= 
    -\frac{k_1}{2k_2} + \frac{\exp\left(\frac{k_1^2}{4k_2}\right)}{\sqrt{\pi k_2}\erfc \left(\frac{k_1}{2\sqrt{k_2}}\right)} \label{eq:tGaussa} \\
    \langle t\rangle_{\rm FSSH} &= \frac{7}{12}\tau. \label{eq:tGaussb}
\end{align}    
\end{subequations}
[The factor 7/12 arises due to the population gain of the active state. Neglecting the last term in Eq.~\eqref{eq:Dfssh} would instead give 1/2.]
By setting $\langle t\rangle_{\rm Gauss}=\langle t\rangle_{\rm FSSH}$ and solving for $\tau$, we finally obtain the decoherence time
\begin{equation}\label{eq:tauGauss}
    \tau_{\rm Gauss} \equiv \frac{12}{7}\left[-\frac{k_1}{2k_2} + \frac{\exp\left(\frac{k_1^2}{4k_2}\right)}{\sqrt{\pi k_2}\erfc \left(\frac{k_1}{2\sqrt{k_2}}\right)}\right].
\end{equation} 
This particular formula appears (as far as I am aware) to be a new result. However, it is merely a refinement of numerious previous formulas based on Gaussian overlaps, and does not contribute any significant new insight in itself.

The assumptions behind this formula are that the wavepackets have (i) Gaussian shape, (ii) constant width, and (iii) follow trajectories with constant force difference.
Let us check to what extent these assumptions are valid for the wavepacket simulation of Tully's extended coupling model shown in Fig.~\ref{fig:tully3WP}. Already from the snapshots in the two left columns, one can see that the $q$ and $p$ distributions deviate substantially from Gaussian shape during the crossings. To quantify the shape of the evolving wavepackets, we use the moments
\begin{equation}
    \mu_n = \int dq \, (q-q_0)^n |\psi(q)|^2.
\end{equation}
The third column in Fig.~\ref{fig:tully3WP} shows the mean ($\mu_1$), the standard deviation ($\sqrt{\mu_2}\equiv \sigma$), the skewness ($\mu_3/\sigma^3$), and the excess kurtosis ($\mu_4/\sigma^4-3$) for $q$. The fourth column shows the corresponding statistics for $p$. For an ideal Gaussian, both the skewness and excess kurtosis should vanish, and the observation that they do not is a clear indication of the non-Gaussian shape. 

For this reason, it should not come as a surprise that running FSSH with the Gaussian decoherence timescale in Eq.~\eqref{eq:tauGauss} leads to too strong coherence suppression (see Fig.~\ref{fig:tully3}b). 
The ultimate reason for this error is that $D_\text{Gauss}(t)$ decays much faster than the overlap function of the true quantum wavepacket.

\subsection{Nonadiabaticity threshold}\label{sec:threshold}
If the potentials $V_a$ and $V_b$ had been globally harmonic, the Gaussian approximation would have been exact. Unfortunately, almost all regions of strong nonadiabatic coupling entail potentials that are undeniably anharmonic. Despite strenuous efforts since many decades, it is still practically impossible to reliably calculate the decoherence rate in such regions.

A pragmatic way to deal with this problem is to \emph{not apply any decoherence correction at all} unless the nonadiabatic coupling is sufficiently weak.
The error of applying too little decoherence is certainly preferable to applying too much decoherence, since in the latter case we face the risk that the ``correction'' destroys the correct quantum evolution of the coherences. 

Once the trajectory has left the region of nonadiabatic coupling, it is again in a potential that we can make more accurate statements about. For almost all cases of interest, the uncoupled potentials can be well approximated as either harmonic or linear. In harmonic models, it is well-justified to calculate $\tau_{\rm Gauss}$ from Eq.~\eqref{eq:tauGauss}. 
We continue to neglect wave packet broadening for the sake of simplicity, meaning that $\gamma$ remains the same as in the initial condition.
Tully's model is not harmonic, but only involves two crossings in immediate succession, so wave packet broadening is expected to be of minor importance. 

\begin{figure}
    \centering
    \resizebox{0.7\columnwidth}{!}{\includegraphics{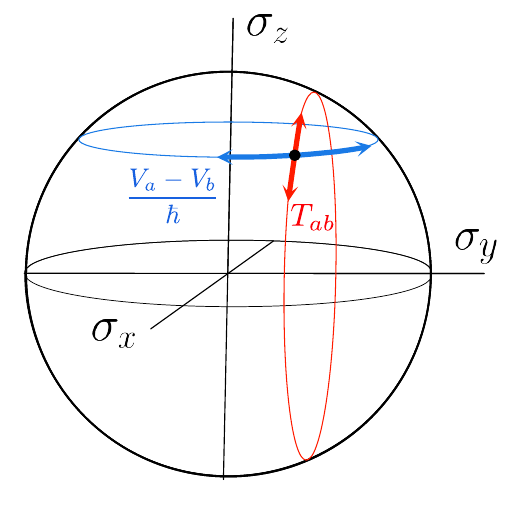}}
    \caption{Bloch sphere with axes $\sigma_x=2\Re c_a^*c_b$, $\sigma_y=2\Im c_a^* c_b$, $\sigma_z=|c_a|^2-|c_b|^2$. The frequency of population dynamics is proportional to $T_{ab}$, and the frequency of coherence dynamics is proportional to $(V_a-V_b)/\hbar$. The ratio of their magnitudes is a dimensionless measure of nonadiabaticity, Eq.~\eqref{eq:r}. (The population rotation axis shown here corresponds to the case where $T_{ab}$ is real.)}
    \label{fig:bloch}
\end{figure}

How then should one identify regions of weak/strong coupling? Fang and Hammes-Schiffer have proposed to reset the amplitudes when the magnitude of the NACV becomes smaller than a preset tolerance.\cite{Fang1999improvement} Unfortunately, the optimal value for such a tolerance will be strongly system-dependent. 
A more general alternative is to measure nonadiabaticy through the Massey parameter
\begin{equation}\label{eq:r}
    r = \frac{\hbar|T_{ab}|}{|V_a-V_b|}.
\end{equation}
The Massey parameter is dimensionless and has been used in various contexts of semiclassical dynamics.\cite{Nikitin1968,Stine1976,Loaiza2018}
On a Bloch sphere (see Fig.~\ref{fig:bloch}), $T_{ab}$ is proportional to the frequency of rotation around the coherence axis (leading to flow of population), whereas $(V_a-V_b)/\hbar$ is proportional to the frequency of rotation around the $\sigma_z$ axis (not leading to flow of population). 
If $r$ is much smaller than 1, the states are essentially uncoupled, and if $r$ is larger than 1, the states are strongly coupled.

My suggestion is to only apply the decoherence correction in regions where
\begin{equation}\label{eq:r_criterion}
    r<r_0,
\end{equation}
which is the most important equation in this article.
The parameter $r_0$ will be referred to as the nonadiabaticy threshold (NT). This threshold should be small enough to avoid the issues observed in section~\ref{sec:Gaussian} (i.e., much smaller than 1), but not so small that it turns off decoherence completely. It is \emph{a priory} not clear whether a single $r_0$ value will suit for all systems. But by virtue of being dimensionless, it is also not unlikely.
In the following, I will refer to the combination of the decoherence time in Eq.~\eqref{eq:tauGauss} with the criterion $r<r_0$ as the ``Gaussian Overlaps with Nonadiabaticity Threshold'' (GONT) method. Each non-active state has a separate $r$ and $\tau_{\rm Gauss}$.

The impact of the threshold is illustrated for Tully's extended coupling model in Fig.~\ref{fig:tully3}b-c. With a loose threshold of $r_0=0.1$, the GONT method leads to slight improvements in accuracy compared to using no threshold $(r_0\to\infty)$, but is still insufficient. The tighter threshold $r_0=0.01$ may be considered adequate, whereas $r_0=0.001$ leads to excellent accuracy for the populations as well as for the coherence. 
The results remain of similar quality for $r_0=10^{-4}$, albeit with a slightly larger inconsistency between the two population measures. For even smaller thresholds (not shown), the results gradually approach the decoherence-free results.
Hence, the initial assessment based on this single example is that there exists a window of ``safe'' thresholds between $10^{-4}$ and (roughly) 0.01. This window will be narrowed down in section~\ref{sec:results} by testing the method for a wider range of systems.

The reader is invited to compare these results to previous simulations of the same model using more sophisticated coupled-trajectory dynamics based on exact factorization (XF)\cite{min2015EF,agostini2016coupled} and its combination with surface hopping.\cite{VillasecoArribas2022,VillasecoArribas2023energy} Another relevant comparison is the (independent-trajectory) combination of XF with the quantum-trajectory surface hopping (QTSH)\cite{martens2019} known as QTSH-XF.\cite{Dupuy2024} 
In comparison to all these methods, GONT appears to be both simpler and more accurate (at least for this particular example).

Before we continue, it is worth asking whether the NT criterion would be suitable also together with other formulas for the decoherence rate? 
This question is relevant because $\tau_{\rm Gauss}$ requires calculating an additional force per non-active state compared to the decoherence-free simulation, plus a momentum rescaling along the NACV at each time step. In \emph{ab initio} applications, such calculations would likely constitute a computational bottleneck. For this reason, I also suggest a simplified method that replaces $\tau_{\rm Gauss}$ by 0, i.e., collapses a coefficient instaneaneously as soon as $r<r_0$. The wavefunction is renormalized at the end of each time step. Instantaneous collapse has been used in many earlier works\cite{Hammes-Schiffer1994,Fang1999improvement,BedardHearn2005}, but the criterion $r<r_0$ is likely to be more generally applicable than those previously considered. I will refer to this simpler approach as the ``Instantaneous Nonadiabaticity Threshold'' (INT) method. Note that calculating $r$ requires neither forces nor NACVs [since one can compute $T_{ab}$ from Eq.~\eqref{eq:Tab}] so the INT method adds no computational overhead to ordinary FSSH. 

For Tully's model, the INT method leads to slightly less accurate results (shown in Fig.~\ref{fig:tully3}d) than the GONT method  regardless of the value of the threshold. Hence, the rest of the article will focus mainly on the GONT method, but occasionally refer to results of the INT method shown in the supplementary information.

[As a side note, Bai \emph{et al.} have recently proposed another decoherence restriction.\cite{Bai2018trivial} Their method uses the population of the non-active state as the threshold parameter and is intended to prevent collapsing after erroneous hops during unavoided crossings, which is a different problem than the one addressed here.]

\subsection{Energy-based decoherence correction}\label{sec:EDC}
To assess the performance of the GONT decoherence correction, its results will be compared to one of the most commonly used decoherence schemes, the energy-based decoherence correction (EDC) by Granucci and Persico.\cite{granucci2007} The EDC originated as an approximation to previous schemes by Truhlar and co-workers.\cite{Zhu2004csdm,Zhu2005jtct} 
Just like in the Gaussian overlap method of section~\ref{sec:Gaussian}, the EDC dampens the amplitudes of the non-active states and adjusts the active state amplitude such that total population is conserved:
\begin{subequations}
\begin{align}\label{eq:expdamp}
    c'_b &= c_b e^{-\Delta t/\tau_{\rm EDC}} \\
    c'_a &= \frac{c_a}{|c_a|} \left(1-\sum_{b\neq a}|c_b|^2\right)^{1/2}.
\end{align}
\end{subequations}
The only difference is in the damping timescale, which in the EDC is set to
\begin{equation}\label{eq:tauEDC}
    \tau_{\rm EDC} = \frac{\hbar}{|V_a-V_b|}\left(1+\frac{E_0}{E_\text{kin}}\right),
\end{equation}
where $E_{\rm kin}=\sum_j \frac{p_j^2}{2m_j}$ is the total kinetic energy of the system, and $E_0$ is an empirical constant that is almost always set to $E_0=\SI{0.1}{\au}\approx\SI{2.7}{eV}$. (This value stems from what Truhlar and co-workers suggested to use in their original formula, which involved the kinetic energy of a particular momentum projection.\cite{Zhu2004csdm} However, there is no reason for the same value to be optimal in both formulas.)

This ``correction'' is problematic for multiple reasons: (i) The total kinetic energy is an irrelevant quantity. If one adds any number of uncoupled degrees of freedom, it will change the results of the EDC -- even though such DOFs cannot affect the nonadiabatic process. (The same critique can be raised also against recent adaptations of the EDC with machine-learned parameters.\cite{Shao2023})
To resolve this issue, it has been suggested\cite{Plasser2019rhenium} to either use the kinetic energy of a single degree of freedom or divide the kinetic energy by the number of DOFs. But neither of these options would suffice, because: (ii) the EDC assumes a given energy scale. Consider two systems A and B, where all energy scales of B are ten times the energy scales of A. These two systems should have identical dynamics (with time units for B scaled by 1/10 compared to A). Hence, the EDC is fundamentally incorrect in using a fixed energy scale for $E_0$. (iii) The EDC can easily lead to too strong decoherence, even in situations when the correct dynamics is coherent. This last statement will be demonstrated explicitly in section~\ref{sec:results}. 

Despite these objections, the EDC remains heavily in use today.\cite{Papineau2024,Gomez2024,Xu2024zhenggang,Mannouch2024initial,Farkhutdinova2025} It is simpler than almost all available alternatives and adds no computational effort compared to FSSH without decoherence. And it is -- unfortunately -- found to work reasonably well in many cases (see results for the example of Tully's model in Fig.~\ref{fig:tully3}a).
The reason to use the word ``unfortunately'' is that such results give rise to a deceptive impression of validity.

\section{Results and discussion}\label{sec:results}
The main question to address is whether there exists a value of the threshold $r_0$ that is adequate for a wide variety of systems. To answer this question, the GONT method was tested for the following types of models: (i) scattering probabilities in Tully's models for a range of initial momenta, (ii) the spin-boson model in various parameter regimes, (iii) a Frenkel-exciton model of the Fenna--Matthew--Olson complex, (iv) conical-intersection models, and (v) models involving spin-orbit coupling. This suite of systems is by no means universal, but should broadly represent the standard target applications for a method like FSSH.

Before we continue, it appears necessary to comment on how to choose $\gamma$. In several previous works based on Gaussian overlaps,\cite{Schwartz1996decoherence,Granucci2010,tempelaar2018fssh} $\gamma$ was treated as a free parameter. However, for all the systems considered here, the width parameter is arguably not free.
For Tully's models, $\gamma$ is already defined by the initial condition.
For photochemistry scenarios that begin with vertical excitation of a system in its vibrational ground state, $\gamma$ is also unambiguous: with a ground-state potential of the form $\frac{1}{2}m\omega^2q^2$, the corresponding width parameter is $\gamma=m\omega/\hbar$. Its multi-dimensional version is $\gamma_j=m_j\omega_j/\hbar$ in normal-mode coordinates. 
If, instead, the system starts from a thermal distribution, previous authors have suggested using the high-temperature limit of the width of this distribution.\cite{Neria1993,Schwartz1996decoherence}
However, a thermal density does not correspond to a coherent wave packet. For this reason, we shall not include temperature in the width, but instead represent each trajectory with the same characteristic width as the ground state ($\gamma=m\omega/\hbar$) and let the centres of the wave packets follow a thermal distribution. 
Thus, $\gamma$ is well-defined for all the cases considered in this paper, and not a free parameter. 

Information about convergence parameters (time step and number of trajectories) is listed for each system in Table~\ref{tab:convergence}. Each average over $10^4$ to $10^5$ trajectories took less than 30 minutes on a 10-core laptop for the largest models (and less than one minute for the Tully models and the two-state conical intersection models). 

\begin{table}
    \centering
    \caption{Convergence parameters.}
    \label{tab:convergence}
    \begin{ruledtabular}
    \begin{tabular}{ccc}
        Model  &  Time step & \#Trajectories \\  \hline
        Tully models & 0.2 fs & 100'000 \\ 
        Spin-boson models 1,3, and 4 & 0.1 $\hbar/\Delta$ & 100'000 \\
        Spin-boson model 2 & 0.2 $\hbar/\Delta$ & 10'000 \\
        FMO & 0.25 fs & 20'000 \\
        LVC, pyrazine, and SOC models & 0.1 fs & 10'000 \\
    \end{tabular}
    \end{ruledtabular}
\end{table}

\begin{figure}
    \centering
    \resizebox{\columnwidth}{!}{\includegraphics{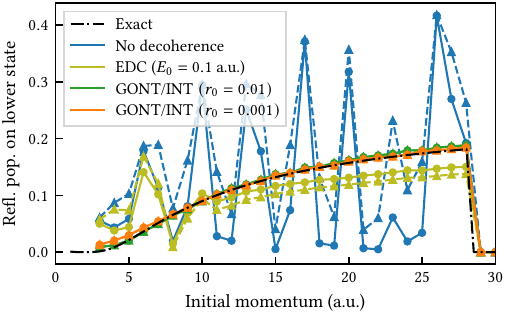}}
    \caption{Asymptotic reflection probability on the lower state in Tully's extended coupling model. 
    Triangles joined by dashed lines indicate wavefunction population and circles joined by solid lines indicate active surface population. Benchmark results using the log-derivative method were taken from Ref.~\onlinecite{runeson2021}.}
    \label{fig:scatt}
\end{figure}

\subsection{Scattering probabilities}
The GONT method was already illustrated for a particular initial condition of Tully's extended coupling model in section~\ref{sec:decoherence}. 
When varying the initial momentum, it is well known that no-decoherence
FSSH produces erroneous fluctuations in the reflected populations (see Fig.~\ref{fig:scatt}). Since this problem is resolved by multiple existing decoherence corrections,\cite{Jain2016efficient,Mannouch2023mash} it is a suitable test scenario for new corrections. 

A subtle difference to the scenario in section~\ref{sec:decoherence} (where $\gamma=0.5$) is that the incoming wave of this test is infinitely narrow in $p$ and infinitely wide in $q$, which formally corresponds to $\gamma\to 0$. Hence, the Gaussian overlap time is zero [since $k_2\to\infty$ in Eq.~\eqref{eq:overlapA}]. For this case, the GONT method is equivalent to the INT method: the amplitude of the non-active state is instantly set to zero whenever $r<r_0$. As shown in Fig.~\ref{fig:scatt}, this simple operation is already sufficient to solve the problem. The reflection probability is accurate no matter if one sets the threshold to $r_0=0.01$ or $r_0=0.001$, and there is no inconsistency between the two population measures.

In contrast, the standard EDC does not remove the oscillations for low initial momenta, and leads to inaccurate results also for large momenta. The fairly accurate results observed in Fig.~\ref{fig:tully3} for the special case $p_0=10$ (albeit with a different width parameter) appears even to be a lucky coincidence.

A side comment is that Tully's original paper did not consider a plane-wave initial condition, but rather a wide wavepacket with width parameter $\gamma=2\left(\frac{p_0}{20\hbar}\right)^2$. In this case, GONT is not equivalent to INT, but nevertheless leads to results that are indistinguishable from those shown in Fig.~\ref{fig:scatt}. (To avoid artificial smoothing, all trajectories start with $p=p_0$ instead of initializing from a Wigner distribution.)

For completeness, the results for Tully's other two models (single and dual avoided crossings) are included in the SI (Figures~\ref{fig:scatt1} and \ref{fig:scatt2}). FSSH describes these models well even without decoherence, so there is nothing left to improve.
The GONT method with $\gamma=2\left(\frac{p_0}{20\hbar}\right)^2$ preserves the no-decoherence accuracy, whereas EDC destroys the St\"{u}ckelberg oscillations in the dual crossing model (due to oversuppression of the coherence). The INT method requires a lower threshold in the dual crossing model, but remains consistent with the GONT method for the single crossing model.

\begin{figure*}
    \includegraphics{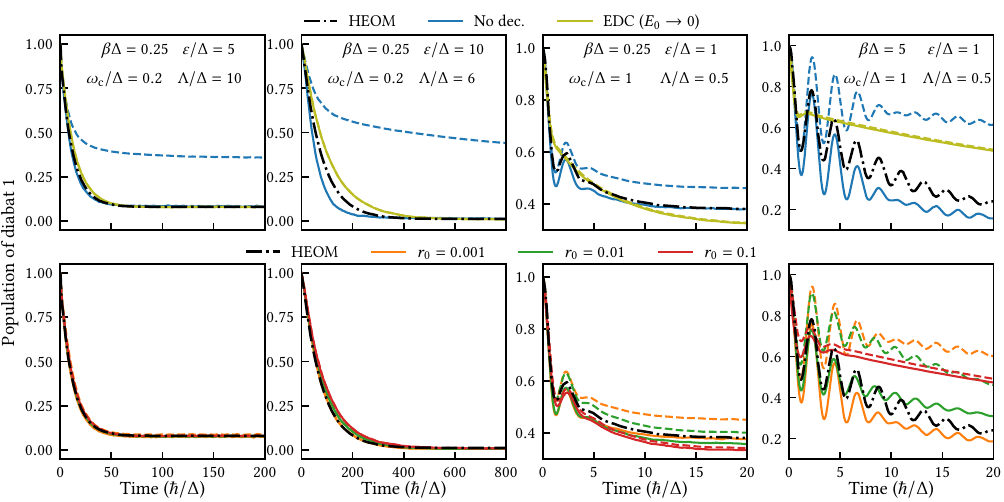}
    \caption{Four cases of the spin-boson model: the activationless regime (column 1), the inverted regime (column 2), the coherent high-temperature regime (column 3), and the coherent low-temperature regime (column 4). The standard population measure $\Pi_1$ (solid lines) is compared to the wavefunction population measure $P_1$ (dashed lines) and to fully quantum HEOM benchmarks\cite{Mannouch2023mash} (black dash-dotted). 
    }\label{fig:spinboson}
\end{figure*}

\subsection{Spin-boson model}
Another paradigmatic system is the spin-boson model
\begin{multline}
    V(q) =\Delta \sigma_x + \left(\varepsilon+\sum_j \kappa_j q_j\right) \sigma_z \\  
    + \sum_j\left( \frac{1}{2}p_j^2 + \frac{\omega_j^2}{2}q_j^2 \right),
\end{multline}
defined with unit masses ($m_j=1$), and with couplings and frequencies specified by the Debye spectral density
\begin{equation}
    J(\omega)=\frac{\pi}{2}\sum_j\frac{\kappa_j^2}{\omega_j}\delta(\omega-\omega_j)=\frac{\Lambda}{2}\frac{\omega\omega_{\rm c}}{\omega^2+\omega_{\rm c}^2}.
\end{equation}
Here, $\Delta$ is the diabatic coupling, $\varepsilon$ (half) the bias, $\Lambda$ the reorganization energy, and $\omega_{\rm c}$ the characteristic frequency of the bath. Together with the thermal energy $k_{\rm B}T$, the model contains five parameters of which only four are independent, since one of them (say, $\Delta$) can always be used as a unit for the others.
We shall not cover the entire range of all parameters but will limit the discussion to four cases that allow for comparison to previous results with other methods (in particular, MASH\cite{Mannouch2023mash} and multi-state MASH\cite{Runeson2023mash}). 
In each case, $q_j$ and $p_j$ were initially sampled from the classical Boltzmann distribution of an uncoupled bath, and the initial electronic density matrix was $|1\rangle\langle 1|=\frac{1}{2}(1+\sigma_z)$. 

The spin-boson model leads to severe formal problems for the EDC. Since the model is defined in reduced units, the constant $E_0=\SI{0.1}{\text{a.u.}}$ is meaningless
(for example, the cases $\Delta=\SI{1000}{\text{a.u.}}$ and $\Delta=\SI{0.001}{\text{a.u.}}$ correspond to the \emph{same} model up to a scaling of the time and other energy units). 
In addition, the kinetic energy depends on the number of discretized bath modes. As the discretization becomes finer, the kinetic energy becomes formally infinite.\cite{Lawrence2024sizeconsistent} For this reason, the EDC will always reach the same result as when setting $E_0=0$. The same conclusion holds for any given value of $\Delta$.
In practice, the simulations used a fixed discretization with 100 modes (which is sufficient to converge the results using a standard mid-point scheme~\cite{Hele2013masters}).

Firstly, we consider the activationless rate regime ($2\varepsilon=\Lambda$) in column 1 of Fig.~\ref{fig:spinboson}. Without decoherence, FSSH gives fairly accurate results when looking at the standard measure ($\Pi_1$), but there is considerable inconsistency with the wavefunction measure ($P_1$). The EDC appears to resolve this problem, as does the GONT method for a wide range of values of the threshold. Secondly, we consider the more challenging inverted regime ($2\varepsilon>\Lambda$) in column 2. Without decoherence, $\Pi_1$ decays too quickly, whereas $P_1$ decays too slowly (and to an incorrect limit). The EDC eliminates the inconsistency, but leads to a population decay that is slightly too slow. The GONT method leads to a more accurate decay rate, in particular for thresholds around $r_0=0.001$.

Thirdly, we consider a coherent regime at high temperature ($\beta\omega_{\rm c}=0.25$) in column 3. Without decoherence, $P_1$ is accurate for short times whereas $\Pi_1$ is better for long times. The EDC does eliminate this inconsistency, but at the price of completely destroying the coherent oscillations. The GONT method solves this problem by limiting decoherence to regions where it is ``safe''. On the other hand, the smaller one makes $r_0$, the smaller the improvement compared to using no decoherence at all. For $r_0=0.001$, the results are almost identical to the no-decoherence method. For $r_0=0.01$, $P_1$ is improved while $\Pi_1$ is slightly worse. For $r_0=0.1$, the inconsistency is almost eliminated, but the long-time limit is too low, so this threshold value is too high.

Finally, we consider a coherent regime at low temperature ($\beta\omega_{\rm c}=5$) in column 4. In this case, one does not expect mixed quantum-classical methods to work well, but it is nevertheless useful to illustrate the breakdown of the method. Without decoherence, FSSH suffers from substantial internal inconsistency. Again, the EDC completely destroys the coherent evolution. In this case, the GONT method is rather sensitive to the value of the threshold. It appears as if $r_0=0.01$ provides a reasonable compromise for $\Pi_1$, but the $P_1$ remains inconsistent.

\begin{figure*}
    \includegraphics{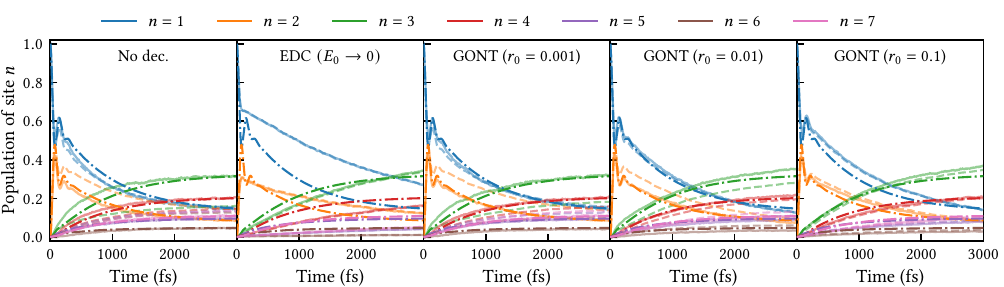}
    \caption{Population dynamics in the FMO model. Each panel compares the $\Pi_n$ population measure (solid lines) to the $P_n$ population measure (dashed) and to fully quantum HEOM benchmarks\cite{Runeson2023mash} (dash-dotted).
    }\label{fig:fmo}
\end{figure*}

These results were obtained using a classical Boltzmann distribution as the initial condition. If one instead starts from an initial Wigner distribution (Figure~\ref{fig:wigner}), the results of the first two columns hardly change. However, the results of the last two columns change significantly and display pronounced overheating for both population measures. Such overheating is expected for the low-temperature case (due to zero-point energy leakage), but is somewhat surprising for the high-temperature case, given that $\beta\omega_{\rm c}=0.25$ is smaller than 1. This comparison reveals that FSSH -- with or without decoherence corrections -- is considerably more sensitive to the change of initial distribution than MASH is.\cite{Mannouch2023mash,Runeson2023mash} As usual, the classical distribution is more accurate for long-time dynamics, whereas the Wigner distribution is better for short-time dynamics.

To conclude the case of the spin-boson model, a reasonable compromise appears to be to use the GONT method with a threshold no larger than $0.01$. In case of remaining inconsistency, the $\Pi_n$ population measure is more reliable than the $P_n$ measure, provided that the nuclei are initialized from a classical Boltzmann distribution.

\subsection{Fenna--Matthews--Olson model}
A test scenario for a system-bath model with more than two states is the Frenkel-exciton model
\begin{multline}
H = H_{\rm s} + \sum_{n,j}\kappa_j q_{j,n}|n\rangle\langle n| + \sum_{n,j}\left(\frac{1}{2}p_{j,n}^2+\frac{\omega_j^2}{2}q_{j,n}^2\right),
\end{multline}
where $n$ runs over the number of sites and $j$ over the number of modes per site.
A standard model of the Fenna--Matthews--Olson (FMO) complex comprises seven sites with identical and independent baths. In the site basis, and in units of cm$^{-1}$, the system Hamiltonian $H_{\rm s}$ is \cite{adolphs2006fmo}
\begin{equation}\label{eq:HS}
    \begin{pmatrix} 
    200 & -87.7 & 5.5 & -5.9 & 6.7 & -13.7 & -9.9 \\ 
    -87.7 & 320 & 30.8 & 8.2 & 0.7 & 11.8 & 4.3 \\
    5.5 & 30.8 & 0 & -53.5 & -2.2 & -9.6 & 6.0 \\
    -5.9 & 8.2 & -53.5 & 110 & -70.7 & -17.0 & -63.3 \\
    6.7 & 0.7 & -2.2 & -70.7 & 270 & 81.1 & -1.3 \\
    -13.7 & 11.8 & -9.6 & -17.0 & 81.1 & 420 & 39.7 \\
    -9.9 & 4.3 & 6.0 & -63.3 & -1.3 & 39.7 & 230
    \end{pmatrix}.
\end{equation}
The spectral density of each site,
\begin{equation}
    J(\omega)= \frac{\pi}{2}\sum_j \frac{\kappa_j^2}{\omega_j}\delta(\omega-\omega_j)= 2\lambda \frac{\omega\omega_{\rm c}}{\omega^2+\omega_{\rm c}^2}
\end{equation}
has reorganization energy $\lambda=\SI{35}{cm^{-1}}$ and characteristic phonon frequency $\hbar\omega_{\rm c} = \SI{106.14}{cm^{-1}}$ ($\omega_{\rm c}^{-1} = \SI{50}{fs}$), and was discretized into 70 modes per site. The bath was initialized from an uncoupled classical Boltzmann distribution, and the electronic state was initialized in site 1.

Contrary to what has been stated elsewhere\cite{Runeson2023mash}, FSSH is actually fairly accurate for the FMO model\cite{Lawrence2024sizeconsistent}, at least for the $\Pi_n$ populations (see Fig.~\ref{fig:fmo}). (Again, the accurate long-time limit necessitates a classical as opposed to a Wigner distribution.)
However, the $P_n$ populations are ``overheated'', analogous to what has been observed with Ehrenfest dynamics.\cite{runeson2020} 
When including decoherence, the results are reminiscent of the spin-boson model in the coherent high-temperature regime. Once again, the EDC completely destroys the coherent oscillations. The GONT method improves consistency between the population measures at the expense of slightly less accurate long-time limits. Overall, the thresholds $r_0=0.001$ and $r_0=0.01$ lead to more accurate $\Pi_n$ populations than $r_0=0.1$.

Regardless of the threshold, the INT method was found to be considerably less accurate than the GONT method, as shown in Fig.~\ref{fig:fmo-inst}.

\begin{figure*}
    \includegraphics{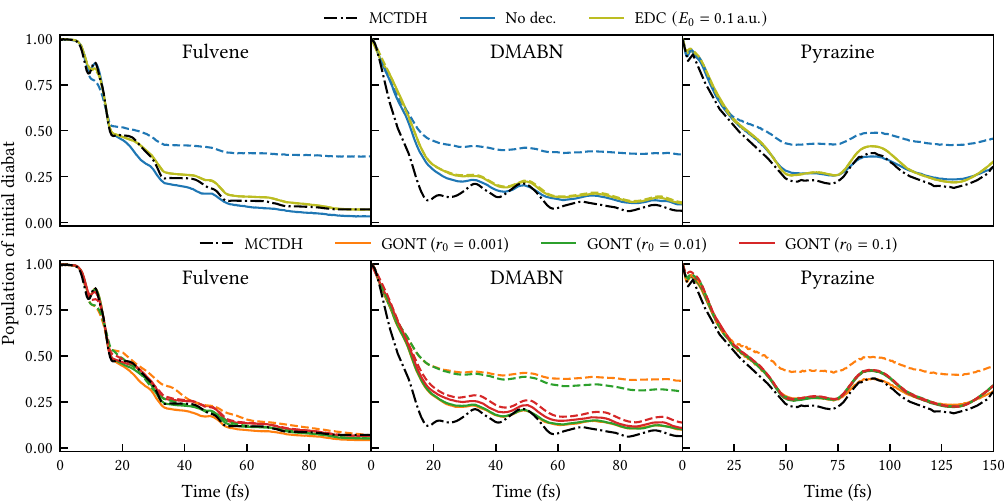}
    \caption{Results for the conical intersection models. FSSH populations are measured using $\Pi_n$ (solid lines) and $P_n$ (dashed lines). 
    }\label{fig:lvc}
\end{figure*}

\subsection{Conical intersection models}
Next, we consider linear vibronic coupling models of the type
\begin{multline}
H_{\rm LVC} = \sum_j \left(\frac{p_j^2}{2m_j} + \frac{1}{2}m_j\omega_j^2q_j^2\right) + \varepsilon_n|n\rangle\langle n| \\
 + \sum_{n}\sum_j \kappa_{n,j}q_j |n\rangle\langle n| + \sum_{n\neq m}\sum_j \lambda_{nm,j} q_j |n\rangle\langle m| 
\end{multline}
The off-diagonal linear coupling $(\lambda_{nm,j})$ gives rise to conical intersections, which distinguishes this model from the spin-boson model.

Recently, Ibele and Curchod have suggested the ``molecular Tully models'' ethylene, fulvene, and 4-(N,N-Dimethylamino)benzonitrile (DMABN) for benchmarking nonadiabatic dynamics.\cite{ibele2020tully} G\'{o}mez \emph{et al.} have constructed LVC models for these three systems and calculated exact quantum dynamics using MCTDH.\cite{Gomez2024} The case of ethylene turned out to involve no electronic population transfer and will not be considered here. Instead, we consider the classic benchmark system pyrazine. The pyrazine model considered here comprises all 24 modes and has a bilinear potential of the type
\begin{multline}
    H = H_{\rm LCV} + \sum_n \sum_{ij} \gamma_{n,ij} q_i q_j |n\rangle\langle n| \\ 
    + \sum_{n\neq m} \sum_{ij} \mu_{nm,ij} q_i q_j|n\rangle\langle m|.
\end{multline}
The model parameters and benchmarks of the fulvene and DMABN models were taken from Ref.~\onlinecite{Gomez2024}. The pyrazine model is fully specified in Ref.~\onlinecite{raab1999pyrazine} and its benchmark is from Ref.~\onlinecite{raab1999density}. The bilinear couplings lead to state-dependent shifts in the nuclear frequencies, but these shifts are so small that there is no reason to consider a state-dependent $\gamma$ when computing the decoherence time from Gaussian overlaps. In all cases, the nuclei were sampled from a ground-state Wigner distribution $(T=0)$ and the electronic state was initialized on the diabat with the highest energy (at ground-state geometry).

The results for all systems are shown in Fig.~\ref{fig:lvc}. Generally, FSSH without decoherence already gives fairly accurate $\Pi_n$ populations. (To obtain this result, it is essential to use momentum rescaling along the NACV, as was recently demonstrated for fulvene and DMABN.\cite{Mannouch2024kelly}) For all three systems, there is considerable inconsistency with the wavefunction ($P_n$) population measure.  
With the GONT method, this inconsistency can be completely removed for fulvene. In this case, the results are not sensitive to the threshold, although $r_0=0.01$ appears to be particularly accurate. For DMABN, the inconsistency remains unless the threshold is increased to $r_0=0.1$, but the $\Pi_n$ measure is robust upon changing the threshold. The comparatively low accuracy of FSSH for the first 20 fs appears to be unrelated to the issue of decoherence. 
For pyrazine, $r_0=0.01$ is sufficient to eliminate inconsistency. Again, $\Pi_n$ is insensitive to the value of the threshold. 

Since the threshold $r_0=0.1$ has already been ruled out by previous tests, the best overall compromise appears to be $r_0=0.01$. The remaining inconsistency for DMABN is acceptable, given the robust results for the $\Pi_n$ populations.

\begin{figure*}
    \includegraphics{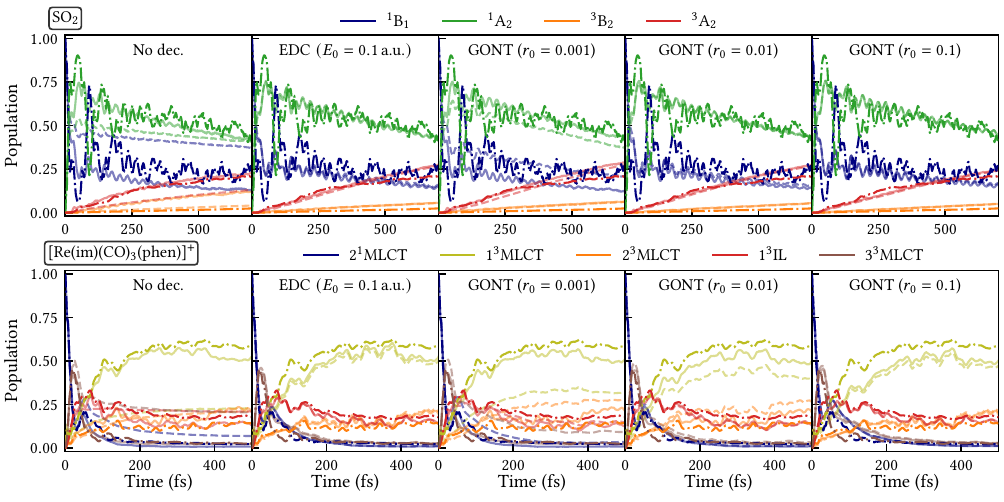}
    \caption{Results for models with spin-orbit coupling. Populations have been summed over each multiplet. Each panel compares $\Pi_n$ populations (solid) to $P_n$ populations (dashed) and to MCTDH benchmarks (dash-dotted).} \label{fig:soc}
\end{figure*}

Interestingly, the EDC is also fairly accurate for all three cases. One should be cautions, however, to conclude that the EDC is safer for conical-intersection models than for system-bath models. There is no clear reason for why the decoherence rate in Eq.~\eqref{eq:tauEDC} would be more justified in either case. It is also not clear how the EDC would perform for an intermediate case with both coherent regions and conical intersections. Perhaps a better explanation for its accuracy in Fig.~\ref{fig:lvc} is that fulvene, DMABN, and pyrazine comprise localized crossings with fast decoherence, so that one can use almost any correction. In fact, for all three models, the simpler INT method (instantaneous decoherence as soon as $r<r_0$) is at least as accurate as the GONT method or the EDC, as shown in Fig.~\ref{fig:lvc-inst}.

\subsection{Spin-orbit coupling}
Finally, we consider LVC models including spin-orbit coupling (SOC)
\begin{equation}
    H = H_{\rm LVC} + H_{\rm SOC}.
\end{equation}
Since the SOC is typically much smaller than the remaining interactions, such models will test whether the GONT method is suitable across multiple energy scales. We consider two models: (i) a model for SO$_2$ constructed by Gonz\'{a}lez and co-workers\cite{Plasser2019lvc,Plasser2019rhenium} at the MR-CISD level of electronic structure, and (ii) a model of the rhenium complex [Re(im)(CO)$_3$(phen)]$^+$ constructed by Fumanal \emph{et al.}\cite{Fumanal2018rhenium}
Gonz\'{a}lez and co-workers have previously produced MCTDH benchmarks for these two models and compared them to FSSH with the EDC,\cite{Plasser2019lvc} and in the case of the rhenium complex also to A-FSSH.\cite{Plasser2019rhenium}
All model parameters (including the details of $H_{\rm SOC}$) are available in Refs.~\onlinecite{Plasser2019lvc} and~\onlinecite{Plasser2019rhenium}.

When running surface hopping on systems with SOC, it is not immediately clear whether it is better to hop between eigenstates of the SOC-free Hamiltonian (the ``spin-diabatic'' approach) or hop between eigenstates of the total Hamiltonian (the ``spin-adiabatic'' approach).\cite{Granucci2012soc} 
The present work used the spin-adiabatic approach because it treats all terms on the same footing. This choice is consistent with previous FSSH simulations for the considered systems.\cite{Plasser2019lvc,Plasser2014}

The only additional difficulty compared to the conical intersection models is that the SOC can be complex, leading to complex eigenstates and NACVs (here, only for the rhenium model; the SO$_2$ model only contains real SOC terms). What is the appropriate direction of momentum rescaling in this case? To answer this question, it is instructive to write the numerator in Eq.~\eqref{eq:Pbdot_over_Pa} (which expresses the flow of population from $a$ to $b$) as
\begin{equation}
    2\Re[c_a^* d_{ab}c_b]\cdot \dot{q}.
\end{equation}
Following the argument in section~\ref{sec:rescaling}, the only projection of the velocity that is involved in the population flow is along the direction of $\Re[c_a^* d_{ab}c_b]$. Hence, this is the more general direction we want to rescale along. Note that this choice differs from Ref.~\onlinecite{Plasser2019rhenium} (which used $\Re d_{ab}$) and Ref.~\onlinecite{Plasser2019lvc} (which, for the present model, used rescaling of the entire momentum). It also differs from other choices considered by Subotnik and co-workers.\cite{Miao2019}

The results for both models are shown in Fig.~\ref{fig:soc}. 
Following the same pattern as for the other conical intersection models, no-decoherence FSSH already gives fairly accurate $\Pi_n$ populations but inaccurate $P_n$ populations. The GONT method with $r_0=0.01$ eliminates the inconsistency for SO$_2$ but not for the rhenium complex. However, a higher threshold of $r_0=0.1$ leads to less accurate $\Pi_n$ populations for the rhenium complex. Thus, we again reach the conclusion that it is safer to use a lower threshold. The difference in $\Pi_n$ between $r_0=0.001$ and $r_0=0.01$ is negligible for SO$_2$ and small (but not negligible) for the rhenium complex. 

For the rhenium complex, the EDC results presented here differ slightly from Ref.~\onlinecite{Plasser2019rhenium} because of the different momentum rescaling direction and because the previous calculations only used 200 trajectories. Nevertheless, the qualitative behaviour is the same. Incidentally, the A-FSSH results reported in Ref.~\onlinecite{Plasser2019rhenium} are qualitatively similar to the GONT results reported here with $r_0=0.001$, i.e., the decoherence is considerably weaker with A-FSSH compared to the EDC. The same observation has made also for other comparable systems.\cite{Heller2021decoherence} 

With the simpler INT method and a $r_0=0.001$ threshold, the results shown in Fig.~\ref{fig:soc-inst} are actually as good as or better than all the other decoherence corrections. The larger threshold $r_0=0.01$ is slightly worse, which indicates that INT is a more invasive method than GONT and, hence, requires a lower threshold.

\section{Concluding remarks}\label{sec:conclusions}
It appears to be near practically impossible to calculate reliable decoherence rates in anharmonic systems. For this reason, \emph{ad hoc} decoherence corrections have a risk to harm more than they help. 
A pragmatic but general solution to this problem is to restrict decoherence corrections to regions of low nonadiabaticity, as measured by the dimensionless Massey parameter. Extensive tests have been run to identify suitable values of the threshold parameter $r_0$.

In practice, I suggest running two simulations: one with $r_0=0.01$ and another with $r_0=0.001$. Then $\Pi_n$ averaged over the two simulations is a robust population measure for state $n$, and the \emph{difference} between their $\Pi_n$ indicates its uncertainty. Such a procedure yields accurate results for the entire suite of test systems, and it correctly identifies the most sensitive cases (the low-temperature spin-boson model, FMO, and the rhenium complex). The $P_n$ populations will sometimes remain inconsistent with $\Pi_n$, but this outcome is acceptable when the $\Pi_n$ population is stable (and preferable to a simulation with too strong decoherence).

One may broadly classify the studied systems by whether their decoherence time is long or short compared to the duration of the population transfer. The first type (slow decoherence) includes the coherent spin-boson models and FMO (and, more generally, systems with substantial recrossing). This type is particularly precarious, and unrestricted decoherence corrections are likely to destroy the interesting part of their dynamics. The second type (fast decoherence) includes the incoherent spin-boson models and the conical intersection models without SOC. This type is fairly insensitive to the choice of decoherence correction. The simplest approach is to use instantaneous decoherence when $r<r_0$, which adds no computational overhead compared to ordinary FSSH and is at least as reliable as more complicated decoherence corrections. However, this simpler approach does \emph{not} work for the slow-decoherence type of systems (and only qualitatively for models with SOC, which may be regarded as an intermediate type). The Gaussian overlap decoherence time in Eq.~\eqref{eq:tauGauss}, on the other hand, was found to work reasonably well for the entire set of test systems and is therefore expected to be suitable also in the general case.

A practical limitation of the Gaussian overlap formula is that it requiries additional force calculations and momentum rescalings at each time step.
However, thanks to the rapid development of machine-learned potentials, such calculations are expected to become cheaper than they are currently for \emph{ab initio} systems. Another limitation is that the width parameter matrix $\gamma$ may be difficult to define for atomistic systems that lack a clear separation into normal modes. In such cases, one could still combine the nonadiabaticity threshold with some other decoherence formula. However, there is no reason to combine the NT with the EDC, because such a method would not resolve the objections listed in section~\ref{sec:EDC}. 
A final problem of the Gaussian overlap formula is that the NACV between uncoupled surfaces may be exactly zero, which would lead to undefined $\Delta p_{\rm hop}$. However, such cases are clearly of the ``fast decoherence'' type considered above, for which the appropriate decoherence time is zero (a possible solution is discussed in Fig.~\ref{fig:scatt1}). 

Could one use the same decoherence scheme also for MASH? Due to the different role of the ``wavefunction'' -- which in MASH is merely a point of a phase-space distribution -- this question cannot be fully addressed within the scope of the present paper. Collapsing onto a pole of the Bloch sphere would not lead to the desired behaviour. 
Simple decoherence corrections do already exist for the original two-state MASH,\cite{Mannouch2023mash} for example by resampling the distribution to match the active adiabat when the energy gap is large compared to $\kBT$.\cite{Lawrence2023mash} The nonadiabaticity threshold should provide a more general criterion to trigger such resamplings. For the multi-state adaptations of MASH\cite{Runeson2023mash,Lawrence2024sizeconsistent} the problem is much harder to analyse and is left for future work.

\section*{Supplementary material}
The supplementary material includes results for Tully's single and dual avoided crossing models, the spin-boson model initialized with the Wigner distribution, as well as results for the INT method.

\section*{Acknowledgements}
I would like to thank 
Michael Thoss and Gerhard Stock for useful discussions. The research was funded by an Alexander von Humboldt Fellowship.

\section*{Author declarations}
The author has no conflicts of interest to disclose.

\section*{Data availability}
Data presented in the figures are available from the corresponding author upon request. 

\bibliography{runerefs}

\newpage

\appendix
\onecolumngrid

\renewcommand\thefigure{S\arabic{figure}}  
\setcounter{figure}{0} 
\renewcommand\theequation{S\arabic{equation}}  
\setcounter{equation}{0} 
\setcounter{page}{1}

\section*{Supplementary information: Additional results}

\noindent Tully's single avoided crossing model is defined by the diabatic potential
\begin{align}
    V_{11} &= - V_{22} = A \frac{q}{|q|}(1-e^{-B|q|})  \nonumber \\
    V_{12} &= V_{21} = Ce^{-Dq^2} \label{eq:tully1}
\end{align}
with $A=0.01$, $B=1.6$, $C=0.005$, and $m=2000$ in atomic units.

\begin{figure*}[hb]
    \includegraphics{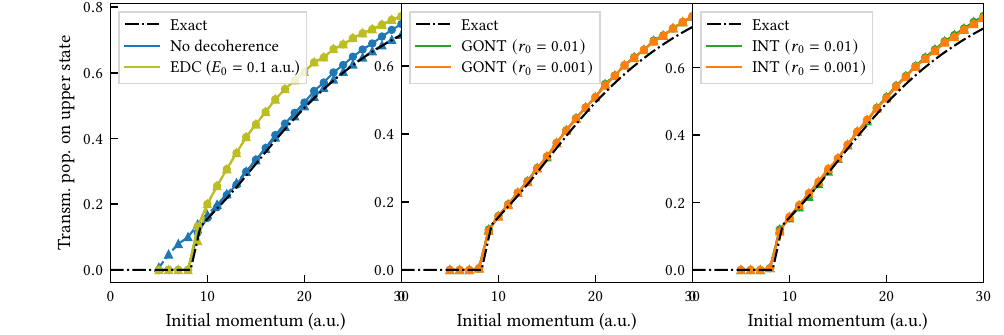}
    \caption{Results for Tully's single avoided crossing model.  Since the NACV was too small to define $\Delta p_{\rm hop}$ (and thereby $\tau_{\rm Gauss}$) in the asymptotic region, the decoherence was taken to be instantaneous in the GONT method whenever $|T_{ab}|\Delta t$ is smaller than machine precision.}\label{fig:scatt1}
\end{figure*}

\noindent
Tully's dual avoided crossing model is defined by the diabatic potential
\begin{align}
    V_{11} &= 0, \quad V_{22} = -A e^{-Bq^2} + E  \nonumber \\
    V_{12} &= V_{21} = Ce^{-Dq^2} \label{eq:tully1}
\end{align}
with $A=0.1$, $B=0.28$, $C=0.015$, $D=0.06$, $E=0.05$, and $m=2000$ in atomic units.



\begin{figure*}[hb]
    \includegraphics{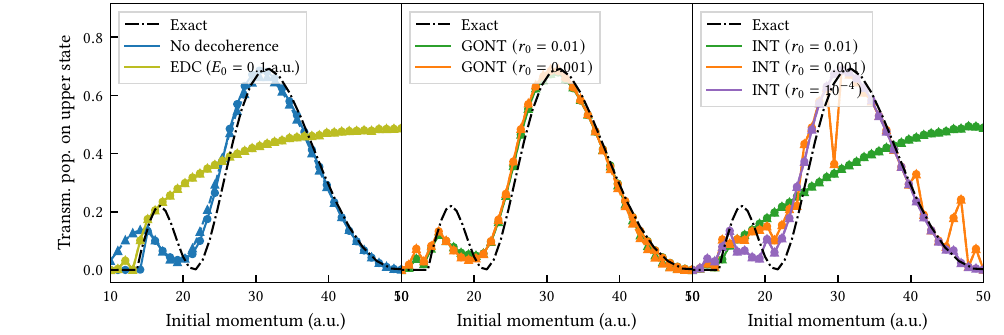}
    \caption{Results for Tully's dual avoided crossing model.}\label{fig:scatt2}
\end{figure*}

\begin{figure*}[hb]
    \includegraphics{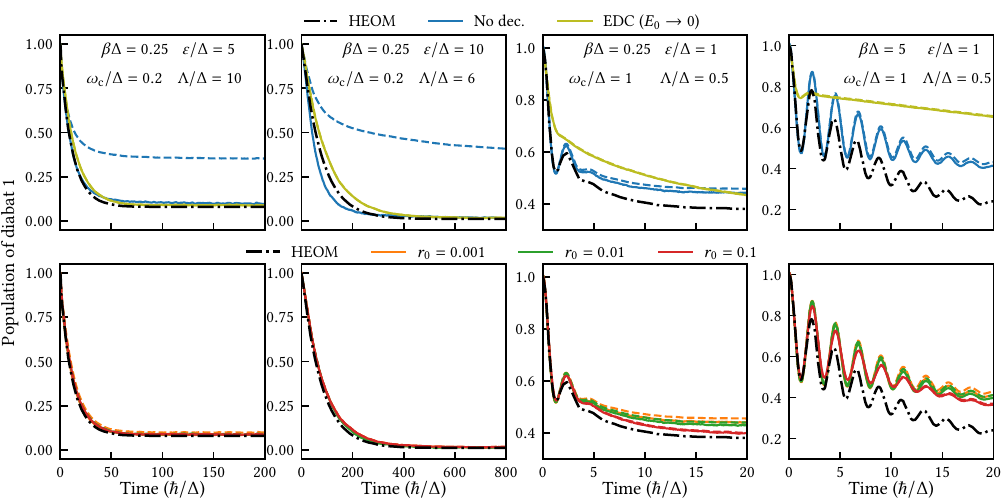}
    \caption{Results for the spin-boson model when initializing the nuclear DOFs from a thermal Wigner distribution $\rho_{\rm wig}(p,q)=\prod_j \frac{\alpha_j}{\pi\hbar} \exp\left[-\frac{2\alpha_j}{\hbar\omega_j}(\frac{p_j^2}{2}+\frac{1}{2}\omega_j^2 q_j^2)\right]$ with $\alpha_j=\tanh(\frac{1}{2}\beta\hbar\omega_j)$.}\label{fig:wigner}
\end{figure*}

\begin{figure*}[hbt]
    \includegraphics{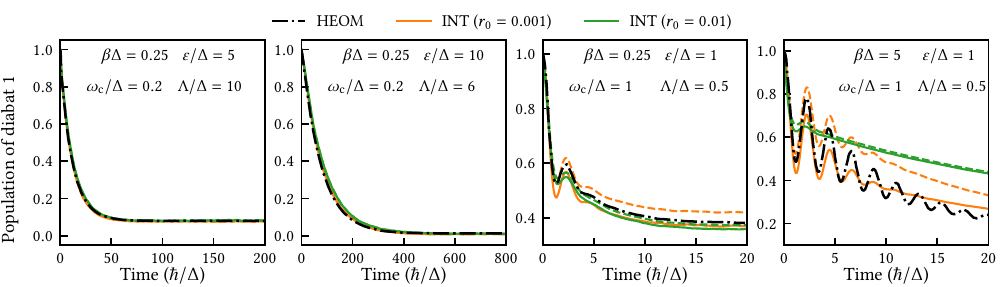}
    \caption{Spin-boson model with instantaneous decoherence when $r<r_0$.}\label{fig:sbm-inst}
\end{figure*}

\begin{figure*}[hbt]
    \includegraphics{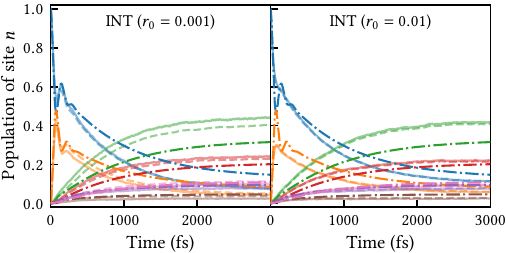}
    \caption{FMO with instantaneous decoherence when $r<r_0$. The colours are the same as in Fig.~\ref{fig:fmo}.}\label{fig:fmo-inst}
\end{figure*}

\begin{figure*}[hbt]
    \includegraphics{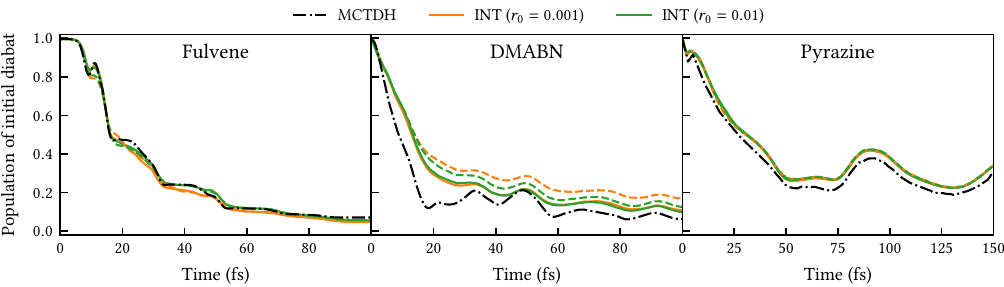}
    \caption{Conical intersection models with instantaneous decoherence when $r<r_0$.}\label{fig:lvc-inst}
\end{figure*}

\begin{figure*}[hbt]
    \includegraphics{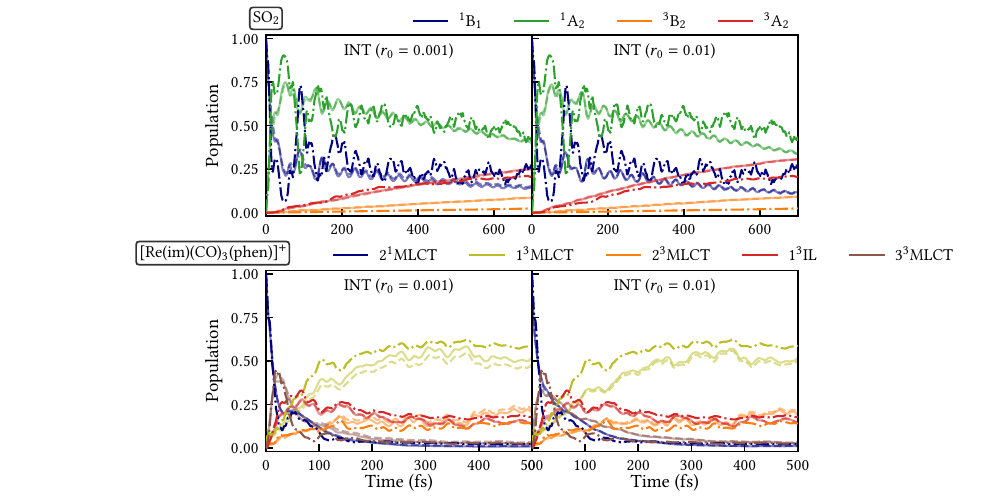}
    \caption{Spin-orbit coupling models with instantaneous decoherence when $r<r_0$. Populations have been summed over each multiplet.}\label{fig:soc-inst}
\end{figure*}

\end{document}